\documentclass[aip,jcp,%
 amsmath,
 amssymb,
 reprint,
 citeautoscript,
 float
]{revtex4-1}
 
\usepackage[utf8]{inputenc}

\usepackage[right=1in,left=1in,top=1in,bottom=1in]{geometry}
\usepackage{mathtools} 
\usepackage{braket}
\usepackage{xcolor}
\usepackage[english]{babel}

\DeclarePairedDelimiter\sqbrak{ [ }{ ] }
\DeclarePairedDelimiter\paran{ ( }{ ) }

\newlength{\CapLen}
\AtBeginDocument{\settoheight{\CapLen}{K}} % after \normalsize
\newcommand{\Kappa}{\resizebox{!}{\CapLen}{$\kappa$}}
\newcommand*{\denM}{D^{\text{M}}}

\newcommand*{\denJ}{D^{\text{J}}}
\newcommand*{\denK}{D^{\text{K}}}
\newcommand*{\denR}{D^{\text{R},n}}
\newcommand*{\azvalvir}{\mathbf{a} \sqbrak*{\mathbf{z}^{\text{val-vir}}} }
\newcommand*{\azvalocc}{\mathbf{a} \sqbrak*{\mathbf{z}^{\text{val-occ}}} }

\newcommand*{\citen}[1]{%
  \begingroup
    \romannumeral-`\x % remove space at the beginning of \setcitestyle
    \setcitestyle{numbers}%
    \cite{#1}%
  \endgroup   
}

\begin{document}

\title{Analytical Gradients for Molecular-Orbital-Based Machine Learning}

\author{Sebastian J. R. Lee}
\affiliation{Division of Chemistry and Chemical Engineering, California Institute of Technology, Pasadena, California 91125, United States}

\author{Tamara Husch}
\affiliation{Division of Chemistry and Chemical Engineering, California Institute of Technology, Pasadena, California 91125, United States}

\author{Feizhi Ding}
\affiliation{%
Entos, Inc.,
%4470 W Sunset Blvd., Suite 107 PMB 94758\\
Los Angeles, CA 90027
}%

\author{Thomas F. Miller III}
\email[]{tfm@caltech.edu}
\affiliation{Division of Chemistry and Chemical Engineering, California Institute of Technology, Pasadena, California 91125, United States}
\affiliation{%
Entos, Inc.,
%4470 W Sunset Blvd., Suite 107 PMB 94758\\
Los Angeles, CA 90027
}%

\date{\today}

\begin{abstract}
Molecular-orbital-based machine learning (MOB-ML) enables the prediction of accurate correlation energies at the cost of obtaining molecular orbitals.
Here, we present the derivation, implementation, and numerical demonstration of MOB-ML analytical nuclear gradients which are formulated in a general Lagrangian framework to enforce orthogonality, localization, and Brillouin constraints on the molecular orbitals.
The MOB-ML gradient framework is general with respect to the regression technique (e.g., Gaussian process regression or neural networks) and the MOB feature design. 
We show that MOB-ML gradients are highly accurate compared to other ML methods on the ISO17 data set while only being trained on energies for hundreds of molecules compared to energies and gradients for hundreds of thousands of molecules for the other ML methods. 
The MOB-ML gradients are also shown to yield accurate optimized structures, at a computational cost for the gradient evaluation that is comparable to Hartree-Fock theory or hybrid DFT. 
\end{abstract}

\maketitle

\section{Introduction}

Analytical nuclear gradients  are the foundation of the quantum chemical elucidation of complex reaction mechanisms via molecular dynamics simulations and minimum-energy and transition-state structure optimization. 
However, the routine calculation of \textit{ab initio} energies and forces with accurate wave function methods is prohibited by their steep cost, e.g.,  coupled-cluster singles, doubles, and perturbative triples [CCSD(T)] scales as\cite{scuseria_comparison_1990} $N^7$ and full configuration interaction scales as\cite{olsen_passing_1990} $N!$ where $N$ is a measure of system size. 
In recent years, machine learning has opened up a new way of mitigating the cost of quantum chemical calculations. \cite{bartok_gaussian_2010,rupp_fast_2012,lavecchia_machine-learning_2015,gawehn_deep_2016,raccuglia_machine-learning-assisted_2016,wei_neural_2016,smith_ani-1_2017,chmiela_machine_2017,kim_virtual_2017,ulissi_address_2017,segler_neural-symbolic_2017,smith_ani-1_2017,schutt_schnet_2017,butler_machine_2018,lubbers_hierarchical_2018,popova_deep_2018,chmiela_towards_2018,smith_less_2018,smith_less_2018,chmiela_towards_2018,s_smith_outsmarting_2018,christensen_operators_2019,unke_physnet_2019,   profitt_shared-weight_2019,christensen_fchl_2020, zaverkin_gaussian_2020, park_accurate_nodate} 
A particular approach that has proven to be highly data efficient and transferable across chemical space is  molecular-orbital-based machine learning (MOB-ML) method \cite{welborn_transferability_2018, cheng_universal_2019, cheng_regression_2019}.
MOB-ML relies on information of local molecular orbitals to predict the pair-wise sum of a post-Hartree--Fock correlation energy at drastically reduced cost. \cite{welborn_transferability_2018, cheng_universal_2019, cheng_regression_2019}

The gradient theory for MOB-ML is comparable to that of non-canonical wave function-based correlation methods, due to factors that include orbital localization and the  non-variational energy expression. 
There exists only a handful of local wave function-based correlation methods for which this effort has been performed. \cite{schutz_analytical_2004, pinski_analytical_2019} 
In this work, we establish a general Lagrangian framework to obtain the analytical nuclear gradients of the MOB-ML energy. 
The framework enforces orthogonality, localization, and Brillouin constraints on the molecular orbitals (section~\ref{sec:mobml_gradient}).
A noteworthy aspect of this framework is that it is agnostic to the training data used for the MOB-ML model, thereby yielding accurate gradient predictions for wave function theory methods for which analytical gradients have not yet been derived or implemented. 
Furthermore, the computational cost of evaluating the MOB-ML energy gradient is comparable to that of a Hartree--Fock (HF) gradient or a hybrid density functional theory (DFT) gradient, such that it is orders of magnitude faster than evaluating the gradients of \textit{ab initio} wave function theories. 

We numerically validate the MOB-ML gradient theory by comparison to energy finite differences in section~\ref{sec:results}.  
Furthermore, we  show that using only training data based on energy calculations (not gradients), MOB-ML efficiently and accurately yields gradients for diverse sets of molecules (section~\ref{sec:results}). 
Comparison of MOB-ML to other ML methods on the example of the ISO17 data set highlights the data efficiency and high transferability of MOB-ML for gradient predictions.
We show that MOB-ML optimized structures for molecules in the ISO17 set are systematically improved with respect to the reference HF method and we compare the performance to that of a standard DFT functional.

\section{MOB-ML Analytical Nuclear Gradients}

\subsection{MOB-ML Energy Theory}
\label{sec:energy}

MOB-ML relies on molecular orbital information from a HF calculation to predict a wave function correlation energy. 
The working equation for the MOB-ML energy is \cite{welborn_transferability_2018, cheng_universal_2019, cheng_regression_2019}
\begin{equation} \label{eq:energy}
    E_{\text{MOB-ML}} \sqbrak*{\mathbf{f}} = E_{\text{corr}} \sqbrak*{\mathbf{f}} + E_{\text{HF}} \text{,}
\end{equation}
where $E_{\text{HF}}$ is the HF energy, and $E_{\text{corr}} \sqbrak{\mathbf{f}}$ is the machine-learned correlation energy,
\begin{equation}
\begin{split}
    E_{\text{corr}} \sqbrak*{\mathbf{f}} & = \sum_{i} \epsilon_{ii} \sqbrak*{\mathbf{f}_{i}} + 2 \sum_{i > j} \epsilon_{ij} \sqbrak*{\mathbf{f}_{ij}} \text{.}
\end{split}
\end{equation}
The matrix of feature vectors, $\mathbf{f}$, is divided into two sub-classes. 
The first sub-class is made up by the diagonal components of $\mathbf{f}$, $\mathbf{f}_{i}$, which represent the valence-occupied orbital $i$.
The second sub-class is made up by the off-diagonal components of $\mathbf{f}$, $\mathbf{f}_{ij}$, which represent the interaction between the valence-occupied orbitals $i$ and $j$. 
Both diagonal and off-diagonal feature vectors are composed of elements from the HF Fock matrix in the MO basis, $\mathbf{F}$, and the MO repulsion integrals, $\boldsymbol\Kappa$, where
\begin{equation}
\begin{split}
\sqbrak*{\boldsymbol\Kappa^{pq}}_{mn} &= ( pq | mn ) \\
    & =  \sum_{\mu \nu \kappa \sigma} C_{\mu p} C_{\nu q} C_{\kappa m} C_{\sigma n} (\mu \nu | \kappa \sigma)  \text{.}
\end{split}
\end{equation}
Here, $(\mu \nu | \kappa \sigma)$ are the four-center atomic orbital integrals with $\mu$, $\nu$, $\kappa$, and $\sigma$ representing atomic orbital indices.
We restrict the MO indices of $\mathbf{F}$ and $\boldsymbol\Kappa$ to the valence-occupied and valence-virtual MOs and we only include 2-center Coulomb- and exchange-type MO integrals, $\sqbrak*{\boldsymbol\Kappa^{pp}}_{qq}$ and $\sqbrak*{\boldsymbol\Kappa^{pq}}_{pq}$ respectively. 
We evaluate the feature vectors following  the protocol that is specified in 
Ref.~\citen{husch_enforcing_2020}.

\subsection{MOB-ML Gradient Theory}
\label{sec:mobml_gradient}

\subsubsection{Lagrangian framework}
\label{sec:mobml_lagrangian}
MOB-ML is a non-variational theory for which the analytical nuclear gradient theory can be derived within a Lagrangian framework, 
\begin{equation}
\begin{split}
     \frac{\text{d} E_{\text{MOB-ML}} }{\text{d} q} = \frac{\text{d} \mathcal{L}}{\text{d} q} &= \frac{ \partial \mathcal{L}}{ \partial q} + \frac{ \partial \mathcal{L}}{ \partial \mathbf{C}} \frac{\partial \mathbf{C}}{ \partial q} \\
    & = \frac{ \partial \mathcal{L}}{ \partial q} \text{,}
\end{split}
\end{equation}
where $q$ refers to nuclear coordinate.
The calculation of the nuclear response of the HF MOs, $\partial \mathbf{C} / \partial q$, is avoided because the Lagrangian $\mathcal{L}$ is minimized with respect to all of its variational parameters which are the MO coefficients, $\mathbf{C}$.
The MOB-ML energy Lagrangian is
\begin{equation}
\label{eq:lagrangian}
\begin{split}
    \mathcal{L} & \sqbrak*{\mathbf{C},\mathbf{x},\mathbf{z},\mathbf{z}^{\text{core}},\mathbf{z}^{\text{val-occ}},\mathbf{z}^{\text{val-vir}}, \boldsymbol{\lambda} } = \\ 
    & E_{\text{MOB-ML}} \sqbrak*{\mathbf{f}} + \sum_{pq} x_{pq} \paran*{ \mathbf{C}^{\dagger} \mathbf{S} \mathbf{C} - \mathbf{I} }_{pq} \\ 
    & + \sum_{ai} z_{ai} F_{ai} \Big|_{i \in \text{occ}, a \in \text{vir}} + \sum_{ri} z_{ri}^{\text{core}} F_{ri} \\
    & + \sum_{i>j} z^{\text{val-occ}}_{ij} r_{ij} + \sum_{ab} z^{\text{val-vir}}_{ab} r_{ab} \\
    & + \sum_{wa} \lambda_{wa} P_{wa} \text{,}
\end{split}
\end{equation}
where $\mathbf{x},\mathbf{z},\mathbf{z}^{\text{core}},\mathbf{z}^{\text{val-occ}},\mathbf{z}^{\text{val-vir}},\text{ and } \boldsymbol{\lambda}$ are the Lagrange multipliers.
We refer to the core MOs with column indices $r, s$, to the valence-occupied localized MOs (LMOs) with column indices $i, j, k, l$, to the valence-virtual LMOs with column indices $a, b$, and to the non-valence-virtual MOs with column indices $w, x$. 
The indices $m, n, p, q$ are used to index generic molecular orbitals. 
The first term on the right hand side (RHS) of Eq.~\ref{eq:lagrangian} is the MOB-ML energy described by Eq.~\ref{eq:energy}. 
The second term on the RHS constrains the HF MOs, $\mathbf{C}$, to be orthonormal, which is commonly referred to as the Pulay force \cite{pulay_ab_1969}. 
The third term on the RHS is known as the Brillouin conditions, which account for the dependence of the correlation energy on the HF optimized molecular orbitals. 
The frozen-core conditions, $F_{ri}=0$, account for neglecting the correlation energy contributions from the core orbitals. 
The localization conditions, $r_{i j} = 0$ and $r_{a b} = 0$, account for how the valence-occupied and valence-virtual MOs are localized respectively. 
In this work, we employ Foster-Boys localization \cite{foster_canonical_1960} and intrinsic bond orbitals (IBO) localization \cite{knizia_intrinsic_2013}, but it is straightforward to generalize to other localization methods.
The valence virtual conditions, $P_{wa} = 0$, reflect how the valence virtual MOs are obtained through a unitary transformation of the virtual MOs. 
This unitary transformation corresponds to the column space of a projection matrix formed by projecting the virtual MOs onto the IAOs. The complementary null space of this projection matrix corresponds to the non valence-virtual orbitals. This projection matrix is defined as
\begin{equation}
    \mathbf{P} = \mathbf{C}^{\text{IAO,}\dagger}_{\text{vir}} \mathbf{C}^{\text{IAO}}_{\text{vir}} \text{,}
\end{equation}
where
\begin{equation} \label{eq:ciao_vir}
    \mathbf{C}^{\text{IAO}}_{\text{vir}} = \mathbf{X}_{\text{occ}}^{\text{IAO}, \dagger} \mathbf{S}_1 \mathbf{C}_{\text{vir}} \text{,}
\end{equation}
and where $\mathbf{C}_{\text{vir}}$ is the virtual MO coefficient matrix.
The matrix $\mathbf{X}^{\text{IAO}}$ transforms between the original AO and IAO basis sets and is expanded in Appendix~\ref{appendix:IBO}.
All together, this yields the following analytical nuclear gradient,
\begin{equation} \label{eq:mobml_grad}
\begin{split}
    \frac{\text{d} E_{\text{MOB-ML}} }{\text{d} q} &=  E_{\text{ML}}^{(q)} + E_{\text{HF}}^{(q)} + \sum_{pq} x_{pq} \paran*{ \mathbf{C}^{\dagger} \mathbf{S}^{(q)} \mathbf{C} }_{pq} \\
    & + \sum_{ai} z_{ai} F_{ai}^{(q)} \Big|_{a \in \text{vir}, i \in \text{occ}} + \sum_{ri} z^{\text{core}}_{ri} F_{ri}^{(q)} \\
    & + \sum_{i>j} z^{\text{val-occ}}_{ij} r_{ij}^{(q)} + \sum_{ab} z^{\text{val-vir}}_{ab} r_{ab}^{(q)} \\
    & + \sum_{wa} \lambda_{wa} P_{wa}^{(q)} \text{,}
\end{split}
\end{equation}
where the superscript $(q)$ denotes the explicit derivative of the quantity with respect to a nuclear coordinate. 
Eq.~\ref{eq:mobml_grad} is the general MOB-ML analytical nuclear gradient and we will now outline how to determine the Lagrange multipliers for our particular use case to arrive at a final working equation. 

\subsubsection{Minimizing 
with respect to MO coefficients}
\label{subsubsec:minimizeL}

All of the Lagrange multipliers ($\mathbf{x}$, $\mathbf{z}$, $\mathbf{z}^{\text{core}}$, $\mathbf{z}^{\text{val-occ}}$, $\mathbf{z}^{\text{val-vir}}$ and $\boldsymbol{\lambda}$) are determined by minimizing the MOB-ML Lagrangian with respect to its variational parameters, which are the MO coefficients, $\mathbf{C}$. Differentiating the Lagrangian with respect to these parameters yields
\begin{equation} \label{eq:stationary_cond}
\begin{split}
    \sum_{\mu} C_{\mu p} &\frac{\partial \mathcal{L}}{\partial C_{\mu q}} = E_{pq} + 2 x_{p q} + \paran*{ \mathbf{D} \sqbrak*{ \mathbf{z} } }_{p q} \\
    & + \paran*{ \mathbf{D} \sqbrak*{ \mathbf{z}^{\text{core}} } }_{p q} + \paran*{ \azvalocc }_{p q} \\ 
    & + \paran*{ \azvalvir }_{p q} + \paran*{ \mathbf{D} \sqbrak*{ \boldsymbol{\lambda} } }_{pq} = 0 \text{,} \\
\end{split}
\end{equation}
where
\begin{equation}
\begin{split}
    E_{pq} & = \sum_{\mu} C_{\mu p} \frac{\partial \paran*{ E_{\text{corr}} \sqbrak*{\mathbf{f}} + E_{\text{HF}} } }{\partial C_{\mu q}}  \\
    & = 4 F_{pq} \Big|_{q \in \text{occ}} + \paran*{ \mathbf{F}  \bar{\mathbf{D}}^{\text{F}} } _{pq} \Big|_{q \in \text{loc}} \\
    & + 2 \paran*{ \mathbf{g} \sqbrak*{\mathbf{C} \bar{\mathbf{D}}^{\text{F}} \mathbf{C}^\dagger }}_{pq} \Big|_{q \in \text{occ}} \\
    & + 2 \sum_{m} \sqbrak*{ \boldsymbol{\Kappa}^{pq} }_{mm} \paran*{ \denJ_{qm} + \denJ_{mq}} \Big|_{mq \in \text{loc}} \\
    & + 2 \sum_{m} \sqbrak*{ \boldsymbol{\Kappa}^{pm} }_{qm} \paran*{ \denK_{qm} + \denK_{mq} } \Big|_{mq \in \text{loc}} \\
    & + \sum_{nm} R^n_{pm} \paran*{\denR_{qm} + \denR_{mq}} \Big|_{mq \in \text{val-occ}} \text{,} \\
\end{split}
\end{equation}
\begin{equation}
\begin{split}
 ( \mathbf{D} &\sqbrak*{\mathbf{z}} )_{pq} = \\
& \quad \sum_{\mu} C_{\mu p} \Bigg( \sum_{a i} z_{ai} \frac{\partial F_{ai}}{\partial C_{\mu q}} \Bigg) \Big|_{i \in \text{occ}, a \in \text{vir}} \\ 
    & = \paran*{ \mathbf{F} \mathbf{z}}_{p q} \Big|_{q \in \text{occ}} \\
    & \quad + \paran*{ \mathbf{F} \mathbf{z}^{\dagger} }_{p q} \Big|_{q \in \text{vir} } + 2 \paran*{ \mathbf{g}[ \bar{\mathbf{z}} ] }_{p q} \Big{|}_{q \in \text{occ}} \text{,} \\
\end{split}
\end{equation}
\begin{equation} \label{eq:dzcore}
\begin{split}
& \paran*{ \mathbf{D} \sqbrak*{\mathbf{z}^{\text{core}}} }_{pq} = \sum_{\mu} C_{\mu p} \Bigg( \sum_{r k} z^{\text{core}}_{rk} \frac{\partial F_{rk}}{\partial C_{\mu q}} \Bigg) \\
    & = \paran*{ \mathbf{F} \mathbf{z}^{\text{core}}}_{p q} \Big|_{q \in \text{val-occ}} \\
    & \quad + \paran*{ \mathbf{F} \mathbf{z}^{\text{core},\dagger} }_{p q} \Big|_{q \in \text{core} } + 2 \paran*{ \mathbf{g}[ \bar{\mathbf{z}}^{\text{core}} ] }_{p q} \Big{|}_{q \in \text{occ}} \text{,}
\end{split}
\end{equation}
\begin{equation}
\begin{split} \label{eq:azloc}
\paran*{ \azvalocc }_{p q} & = \sum_{\mu} C_{\mu p} \Bigg( \sum_{i>j} z_{i j}^{\text{loc}} \frac{\partial r_{i j}}{\partial C_{\mu q}} \Bigg) \text{,} \\
\end{split}
\end{equation}
\begin{equation}
\begin{split} \label{eq:azvalvir}
\paran*{ \azvalvir }_{p q} & = \sum_{\mu} C_{\mu p} \Bigg( \sum_{a>b} z_{a b}^{\text{vir}} \frac{\partial r_{a b}}{\partial C_{\mu q}} \Bigg) \text{,}
\end{split}
\end{equation}
and
\begin{equation} \label{eq:dlambda}
\begin{split}
    ( \mathbf{D} &\sqbrak*{ \boldsymbol{\lambda} } )_{p q} = \sum_{\mu} C_{\mu p} \Bigg( \sum_{wa} \lambda_{wa} \frac{\partial P_{wa}}{\partial C_{\mu q}} \Bigg) \\
    & = ( \mathbf{P} \boldsymbol{\lambda} )_{pq} \Big|_{q \in \text{non-val-vir}} \text{.}
\end{split}
\end{equation}
Eqs.~\ref{eq:azloc} and \ref{eq:azvalvir} are expanded in Appendices \ref{appendix:Boys} and \ref{appendix:IBO}, respectively, $\mathbf{F}$ is the HF Fock matrix, $\mathbf{g}$ includes all of the usual HF two-electron terms, $\mathbf{R}^n$ is expanded in Appendix \ref{appendix:Boys}, the condition $q \in \text{loc}$ restricts the sum to valence-occupied and valence-virtual MOs,  $\bar{\mathbf{z}} = \mathbf{z} + \mathbf{z}^{\dagger}$, $\bar{\mathbf{z}}^{\text{core}} = \mathbf{z}^{\text{core}} + \mathbf{z}^{\text{core,}\dagger}$, and $\bar{\mathbf{D}}^{\text{F}} = \mathbf{D}^{\text{F}} + \mathbf{D}^{\text{F}, \dagger}$.
The matrices $\mathbf{D}^{\text{F}}$, $\mathbf{D}^{\text{J}}$, and $\mathbf{D}^{\text{K}}$ are calculated by
\begin{equation} \label{eq:denM}
\begin{split}
    \denM_{pq} &= \sum_{i} \frac{\partial \epsilon_{ii} \sqbrak*{\mathbf{f}_{i}} }{\partial \mathbf{f}_{i}} \frac{\partial \mathbf{f}_{i}}{\partial M_{pq}} \Big|_{pq \in \text{loc}} \\
    & + 2 \sum_{i > j} \frac{\partial \epsilon_{ij} \sqbrak*{\mathbf{f}_{ij}} }{\partial \mathbf{f}_{ij}} \frac{\partial \mathbf{f}_{ij}}{\partial M_{pq}} \Big|_{pq \in \text{loc}} \text{,} \\
\end{split}
\end{equation}
where $M_{pq}$ refers to $F_{pq}$, $\sqbrak*{\boldsymbol{\Kappa}^{pp}}_{qq}$ and $\sqbrak*{\boldsymbol{\Kappa}^{pq}}_{pq}$, respectively. 
The matrix $\mathbf{D}^{\text{R,n}}$ is
\begin{equation} \label{eq:denR}
\begin{split}
    \denR_{pq} &= 2 \sum_{i > j} \frac{\partial \epsilon_{ij} \sqbrak*{\mathbf{f}_{ij}} }{\partial \mathbf{f}_{ij}} \frac{\partial \mathbf{f}_{ij}}{\partial R^n_{pq}}  \Big|_{pq \in \text{val-occ}} \text{.} \\
\end{split}
\end{equation}
The partial derivatives $\frac{\partial \epsilon_{ii} \sqbrak*{\mathbf{f}_{i}} }{\partial \mathbf{f}_{i}}$ and $\frac{\partial \epsilon_{ij} \sqbrak*{\mathbf{f}_{ij}} }{\partial \mathbf{f}_{ij}}$ on the RHS of Eqns.~\ref{eq:denM} and \ref{eq:denR} are the derivatives of the machine learning prediction with respect to the feature vectors. 

We emphasize that any machine learning method (e.g. Gaussian process regression, regression clustering, neural net, etc.) can be readily used in this gradient framework without modification given $\frac{\partial \epsilon_{ii} \sqbrak*{\mathbf{f}_{i}} }{\partial \mathbf{f}_{i}}$ and $\frac{\partial \epsilon_{ij} \sqbrak*{\mathbf{f}_{ij}} }{\partial \mathbf{f}_{ij}}$. 
Furthermore, we note that the following analytical nuclear gradient derivation generalizes to any type of feature-vector design and construction, so long as the feature-vector elements are obtained from $\mathbf{F}$ and $\boldsymbol{\Kappa}$. 
The partial derivatives $\frac{\partial \mathbf{f}_{i}}{\partial M_{pq}}$, $\frac{\partial \mathbf{f}_{ij}}{\partial M_{pq}}$ and $\frac{\partial \mathbf{f}_{ij}}{\partial R^n_{pq}}$ are expanded in the supplementary material.  

We now proceed to solve for each of the Lagrange multipliers.
First, combining the stationary conditions described by Eq.~\ref{eq:stationary_cond} with the auxiliary conditions $\boldsymbol{x} = \boldsymbol{x}^{\dagger}$ yields the linear Z-vector equations
\begin{equation} \label{eq:LinearZvec}
\begin{split}
\paran*{ 1 - \mathcal{P}_{p q} } &( \mathbf{E} + \mathbf{D} \sqbrak*{\mathbf{z}} + \mathbf{D} \sqbrak*{\mathbf{z}^{\text{core}}} + \azvalocc \\
    & + \azvalvir +  \mathbf{D} \sqbrak*{ \boldsymbol{\lambda} } )_{p q} = 0 \text{,}
\end{split}
\end{equation}
where $ \mathcal{P}_{p q}$ permutes the indices $p$ and $q$, which is used to solve for $\mathbf{z}$, $\mathbf{z}^{\text{core}}$, $\mathbf{z}^{\text{val-occ}}$, $\mathbf{z}^{\text{val-vir}}$ and $\boldsymbol{\lambda}$. 
The matrix $\boldsymbol{x}$ is then obtained as 
\begin{equation} \label{eq:overlap_lag}
\begin{split}
x_{pq} & = - \tfrac{1}{4} \paran*{ 1 + \mathcal{P}_{p q} } \big( \mathbf{E} + \mathbf{D} \sqbrak*{ \mathbf{z}} + \mathbf{D} \sqbrak*{\mathbf{z}^{\text{core}}} \\
& \quad + \azvalocc + \azvalvir + \mathbf{D} \sqbrak*{ \boldsymbol{\lambda}} \big)_{p q} \text{.}
\end{split}
\end{equation} 
The Lagrange multipliers $\mathbf{z}^{\text{val-occ}}$ are solved by considering the (valence-occupied)-(valence-occupied) part of Eq.~\ref{eq:LinearZvec}, yielding
\begin{equation} \label{eq:LinearZvec_occ_occ}
\begin{split}
\paran*{ 1 - \mathcal{P}_{i j} } ( & \mathbf{E} + \mathbf{D} \sqbrak*{ \mathbf{z}} + \mathbf{D} \sqbrak*{\mathbf{z}^{\text{core}}} + \azvalocc \\
 & + \azvalvir + \mathbf{D} \sqbrak*{ \boldsymbol{\lambda}} )_{i j} = 0 \text{.}
\end{split}
\end{equation}
Eq.~\ref{eq:LinearZvec_occ_occ} can be further simplified by showing that
\begin{equation} 
\begin{split}
\paran*{ 1 - \mathcal{P}_{i j} } \paran*{\mathbf{D} \sqbrak*{ \mathbf{z}} }_{i j} & = 0 \text{,} \\
\paran*{ 1 - \mathcal{P}_{i j} } \paran*{\mathbf{D} \sqbrak*{ \mathbf{z}^{\text{core}}} }_{i j} & = 0 \text{,}\\
\paran*{ 1 - \mathcal{P}_{i j} } \paran*{ \azvalvir }_{i j} & = 0 \text{,} \\
\paran*{ 1 - \mathcal{P}_{i j} } \paran*{\mathbf{D} \sqbrak*{ \boldsymbol{\lambda} } }_{i j} &= 0 \text{.} \\
\end{split}
\end{equation}
As a result, $\mathbf{z}^{\text{val-occ}}$ is independent of all other Lagrange multipliers, which simplifies Eq.~\ref{eq:LinearZvec_occ_occ} to
\begin{equation}
\label{eq:zvalocc_cpl}
E_{i j} - E_{j i} + \sum_{k > l} \paran*{ \mathcal{B}_{i j k l} - \mathcal{B}_{j i k l} } z^{\text{val-occ}}_{k l} = 0 \text{,}
\end{equation}
where the 4-dimensional tensor $\mathcal{B}$ is expanded in Appendix \ref{appendix:Boys}.
The set of linear system of equations defined by Eq.~\ref{eq:zvalocc_cpl} are the Z-vector coupled perturbed localization (Z-CPL) equations which are used to solve for $\mathbf{z}^{\text{val-occ}}$.
Subsequently, Eq.~\ref{eq:azloc} can be used to compute $\azvalocc$.

The Lagrange multipliers $\mathbf{z}^{\text{core}}$ are solved by considering the core-(valence-occupied) part of Eq.~\ref{eq:LinearZvec},  yielding
\begin{equation} \label{eq:LinearZvec_core_val}
\begin{split}
\paran*{ 1 - \mathcal{P}_{r i} } ( & \mathbf{E} + \mathbf{D} \sqbrak*{ \mathbf{z}} + \mathbf{D} \sqbrak*{\mathbf{z}^{\text{core}}} + \azvalocc \\
    & + \azvalvir + \mathbf{D} \sqbrak*{\boldsymbol{\lambda}} )_{r i} = 0 \text{,}
\end{split}
\end{equation}
which further simplifies to
\begin{equation} \label{eq:zcore}
\begin{split}
    E_{ri} - E_{ir} + \paran*{ \azvalocc }_{ri} + \paran*{ \mathbf{F} \mathbf{z}^{\text{core}} - \mathbf{z}^{\text{core}} \mathbf{F} }_{r i} = 0 \text{.}
\end{split}
\end{equation}
These are the Z-vector equations used to solve for $\mathbf{z}^{\text{core}}$. Subsequently, Eq.~\ref{eq:dzcore} can be used to calculate $\mathbf{D} \sqbrak*{\mathbf{z}^{\text{core}}}$. 

The Lagrange multipliers $\mathbf{z}^{\text{val-vir}}$ are solved by considering the (valence-virtual)-(valence-virtual) part of Eq.~\ref{eq:LinearZvec}, yielding
\begin{equation} \label{eq:LinearZvec_vir_vir}
\begin{split}
\paran*{ 1 - \mathcal{P}_{a b} } (& \mathbf{E} + \mathbf{D} \sqbrak*{ \mathbf{z}} + \mathbf{D} \sqbrak*{ \mathbf{z}^{\text{core}}} + \azvalocc \\
    & + \azvalvir + \mathbf{D} \sqbrak*{\boldsymbol{\lambda}} )_{a b} = 0 \text{,}
\end{split}
\end{equation}
which further simplifies to
\begin{equation} \label{eq:zcpl_valvir}
E_{a b} - E_{b a} + \sum_{c > d} \mathcal{C}_{a b c d} z^{\text{val-vir}}_{c d} = 0 \text{,}
\end{equation}
where the 4-dimensional tensor $\mathcal{C}$ is expanded in Appendix~\ref{appendix:IBO}.
These are the Z-CPL equations which are used to solve for $\mathbf{z}^{\text{val-vir}}$.
Subsequently, Eq.~\ref{eq:azvalvir} can be used to compute $\azvalvir$.

The Lagrange multipliers $\boldsymbol{\lambda}$ are solved by considering the (non valence-virtual)-(valence-virtual) part of Eq.~\ref{eq:LinearZvec}, yielding
\begin{equation}
\begin{split}
\paran*{ 1 - \mathcal{P}_{w a} } (& \mathbf{E} + \mathbf{D} \sqbrak*{\mathbf{z}} + \mathbf{D} \sqbrak*{\mathbf{z}^{\text{core}}} + \azvalocc \\
    &+ \azvalvir + \mathbf{D} \sqbrak*{\boldsymbol{\lambda}})_{w a} = 0 \text{,} \\
\end{split}
\end{equation}
which further simplifies to
\begin{equation}
\begin{split}
& E_{wa} - E_{aw} + \azvalvir_{wa} - ( \mathbf{P} \boldsymbol{\lambda})_{aw} = 0 \text{.} \\
\end{split}
\end{equation}
These are the Z-vector equations used to solve for $\boldsymbol{\lambda}$. Subsequently, Eq.~\ref{eq:dlambda} can be used to compute $\mathbf{D} \sqbrak*{\boldsymbol{\lambda}}$.

Finally, the Lagrange multipliers $\mathbf{z}$ are solved by considering the virtual-occupied part of Eq.~\ref{eq:LinearZvec}, yielding
\begin{equation}
\begin{split}
( 1 &- \mathcal{P}_{a i} ) ( \mathbf{E} + \mathbf{D} \sqbrak*{\mathbf{z}} + \mathbf{D} \sqbrak*{\mathbf{z}^{\text{core}}} + \azvalocc \\
    &+ \azvalvir + \mathbf{D} \sqbrak*{\boldsymbol{\lambda}} )_{a i} \Big|_{a \in \text{vir}, i \in \text{occ}} = 0 \text{,} \\
\end{split}
\end{equation}
which further simplifies to
\begin{equation} \label{eq:DFTinDFT_External_Occ}
\begin{split}
& E_{ai} - E_{ia} + \paran{ 2 \mathbf{g} \sqbrak*{ \bar{\mathbf{z}}^{\text{core}} }  }_{a i} \\
& \quad + \paran*{ \azvalocc }_{ai} - \paran*{ \azvalvir }_{ia} \\
& \quad - \paran*{\mathbf{P} \boldsymbol{\lambda} }_{i a} + \paran*{ \mathbf{F} \mathbf{z} - \mathbf{z} \mathbf{F} + 2 \mathbf{g} \sqbrak*{ \bar{\mathbf{z}} }  }_{a i} = 0 \text{.} \\
\end{split}
\end{equation}
Here, the MO indices $a$ and $i$ refer to the full virtual and occupied spaces, respectively.
These are the Z-vector coupled perturbed Hartree--Fock (Z-CPHF) equations. 
With the solutions to all Z-vector equations we can return to Eq.~\ref{eq:overlap_lag} to solve for $\boldsymbol{x}$.

\subsubsection{Incorporating molecular-orbital localization} 
\label{sec:mobml_total_gradient}

To provide the working expression of Eq.~\ref{eq:mobml_grad} in terms of derivative AO integrals, we must specify the molecular-orbital localization method. 
For this derivation, we choose the Foster-Boys and IBO localization methods to localize the valence-occupied and valence-virtual orbitals, respectively, such that 
\begin{equation} \label{eq:mobml_grad_ao}
\begin{split}
    \frac{\partial \mathcal{L}}{\partial q} & = \text{tr} \sqbrak*{ \mathbf{d}_{\text{a}} \mathbf{h}^{(q)} } + \text{tr} \sqbrak*{ \mathbf{X}_1 \mathbf{S}_{1}^{(q)} } + \text{tr} \sqbrak*{ \mathbf{X}_2 \mathbf{S}_{2}^{(q)} } \\
    & + \text{tr} \sqbrak*{ \mathbf{X}_{12} \mathbf{S}_{12}^{(q)} } + \sum_{n} \text{tr} \sqbrak*{ \mathbf{W}_n \paran*{\mathbf{R}^{n} }^{(q)} } \\
    & + \tfrac{1}{2} \sum_{\mu \nu \lambda \sigma} D_{\mu \nu \kappa \sigma} (\mu \nu | \kappa \sigma)^{(q)} \text{,} \\
\end{split}
\end{equation}
where $\mathbf{h}$ is the standard one-electron Hamiltonian, $\mu$, $\nu$, $\kappa$ and $\sigma$ label AO basis functions in the original basis, $(\mu \nu| \kappa \sigma)$ are the two-electron repulsion integrals, $\mathbf{S}_2$ is the overlap matrix of the minimal AO basis (MINAO) used in the IBO procedure, and $\mathbf{S}_{12}$ is the overlap matrix between the original AO and MINAO basis sets. 
The effective one-particle density $\mathbf{d}_{\text{a}}$ is defined as
\begin{equation}
    \mathbf{d}_{\text{a}} = \boldsymbol{\gamma} + \tfrac{1}{2}\mathbf{C} \bar{\mathbf{D}}^{\text{F}} \mathbf{C}^\dagger + \tfrac{1}{2} \mathbf{C} \bar{\mathbf{z}} \mathbf{C}^\dagger + \tfrac{1}{2} \mathbf{C} \bar{\mathbf{z}}^{\text{core}} \mathbf{C}^\dagger \text{,}
\end{equation}
where $\boldsymbol{\gamma}$ is the full system HF density. 
The effective two-particle density $\mathbf{D}$ is defined as 
\begin{equation}
\begin{split}
D_{\mu \nu \kappa \sigma} &= \paran*{ \mathbf{d}_{\text{b}} }_{\mu \nu} \gamma_{\kappa \sigma} - \tfrac{1}{2} \paran*{ \mathbf{d}_{\text{b}} }_{\mu \kappa} \gamma_{\nu \sigma} \\
    & + 2 \sum_{pq} \denJ_{pq} C_{\mu p} C_{\nu p} C_{\kappa q} C_{\sigma q} \\
    & + 2 \sum_{pq} \denK_{pq} C_{\mu p} C_{\kappa p} C_{\nu q} C_{\sigma q} \text{,}
\end{split}
\end{equation}
where the effective one-particle density $\mathbf{d}_{\text{b}}$ is defined as
\begin{equation}
    \mathbf{d}_{\text{b}} = \boldsymbol{\gamma} + \mathbf{C} \bar{\mathbf{D}}^{\text{F}} \mathbf{C}^\dagger + \mathbf{C} \bar{\mathbf{z}} \mathbf{C}^\dagger + \mathbf{C} \bar{\mathbf{z}}^{\text{core}} \mathbf{C}^\dagger \text{.}
\end{equation}
The matrices $\mathbf{X}_1$, $\mathbf{X}_2$, $\mathbf{X}_{12}$, and $\mathbf{W}_n$ are defined as 
\begin{equation}
\begin{split} \label{eq:X1}
    \mathbf{X}_1 & = \mathbf{C} \mathbf{x} \mathbf{C}^{\dagger} + \sum_{a > b} \frac{ \partial r_{ab} }{ \partial \mathbf{S}_1 } z_{ab}^{\text{val-vir}}  \\
    & \quad + \sum_{w a} \frac{ \partial P_{wa} }{ \partial \mathbf{S}_1 } \lambda_{wa} \text{,} \\
    \mathbf{X}_2 & = \sum_{a > b} \frac{ \partial r_{ab} }{ \partial \mathbf{S}_2 } z_{ab}^{\text{val-vir}} + \sum_{w a} \frac{ \partial P_{wa} }{ \partial \mathbf{S}_2 } \lambda_{wa} \text{,} \\
    \mathbf{X}_{12} &= \sum_{a > b} \frac{ \partial r_{ab} }{ \partial \mathbf{S}_{12} } z_{ab}^{\text{val-vir}} + \sum_{w a} \frac{ \partial P_{wa} }{ \partial \mathbf{S}_{12} } \lambda_{wa} \text{} \\
\end{split}
\end{equation}
and
\begin{equation} \label{eq:Wxyz}
\begin{split}
    \mathbf{W}_n &= \sum_{i > j} \frac{ \partial r_{ij} }{ \partial \mathbf{R}^n } z_{ij}^{\text{val-occ}} + \mathbf{C} \mathbf{D}^{\text{R},n} \mathbf{C}^\dagger \text{,} 
\end{split}
\end{equation}
where Eq.~\ref{eq:X1} is expanded in Appendix~\ref{appendix:IBO} and Eq.~\ref{eq:Wxyz} is expanded in Appendix~\ref{appendix:Boys}. 

\section{Computational Details}
In this work, we perform calculations on three different data sets: (i) the thermalized water data set published in Ref.~\citenum{cheng_universal_2019}, (ii) a thermalized set of organic molecules featuring up to seven heavy atoms (QM7b-T) \cite{cheng_universal_2019}, and (iii) the ISO17 data set of conformers taken from molecular dynamic (MD) trajectories for constitutional isomers with the chemical formula C$_7$O$_2$H$_{10}$ \cite{schutt_schnet_2017}.

All MOB-ML energy and analytical gradient are implemented in and performed with \textsc{entos qcore}\cite{manby_entos_nodate}.
The DF-HF calculations for the QM7b-T set\cite{cheng_universal_2019}, and the ISO17 set,\cite{schutt_schnet_2017} are performed with a cc-pVTZ \cite{dunning_gaussian_1989} basis set and a cc-pVTZ-JKFIT density fitting basis. \cite{weigend_fully_2002}
The DF-HF calculations for the water calculations are performed with a aug-cc-pVTZ \cite{kendall_electron_1992} and a aug-cc-pVTZ-JKFIT\cite{weigend_fully_2002} basis set.
We employ a molecular orbital convergence threshold of $\texttt{orbital\_grad\_threshold} = 1 \times 10^{-8}$~a.u.
In all MOB-ML calculations, the Foster--Boys \cite{foster_canonical_1960} localization method is used to localize the valence-occupied MOs. 
The valence-virtual space is either localized with Foster--Boys localization (QM7b-T, ISO17) or the IBO localization method \cite{knizia_intrinsic_2013} (water). 
The diagonal and off-diagonal feature vectors are constructed following the procedure outlined in Ref.~\citenum{husch_enforcing_2020}. 
For all Z-CPHF calculations a convergence threshold of $1 \times 10^{-8}$~a.u. is specified.

All WF calculations are performed in Molpro \cite{werner_molpro_2019-1} with the frozen-core approximation, and with density fitting.
All WF pair energy calculations employ the non-canonical MP2 \cite{moller_note_1934,schutz_low-order_1999,hetzer_low-order_2000,werner_fast_2003}
or non-canonical coupled-cluster singles, doubles, and perturbative triples [CCSD(T)] \cite{scuseria_closedshell_1987,scheiner_analytic_1987,scuseria_comparison_1990,lee_analytic_1991,schutz_low-order_2000,schutz_local_2000,werner_efficient_2011} correlation treatments with the cc-pVTZ, cc-pVTZ-MP2FIT, \cite{weigend_efficient_2002}  aug-cc-pVTZ and aug-cc-pVTZ-MP2FIT \cite{weigend_efficient_2002} basis sets. 
An interface between Molpro and \textsc{entos qcore} is used such that WF pair energies are calculated using the DF-HF LMOs produced by \textsc{entos qcore}.
All WF gradient calculations employ the canonical MP2 or 
CCSD(T) correlation treatments with the aug-cc-pVTZ, aug-cc-pVTZ-JKFIT and aug-cc-pVTZ-MP2FIT basis sets. 
For all Z-CPHF calculations needed for the WF gradient an iterative solver with a convergence threshold of $1 \times 10^{-9}$~a.u. is used.

The MOB-ML models for water are trained on non-canonical CCSD(T)/aug-cc-pVTZ pair correlation energies. When constructing the feature vector all non-zero elements from the Fock and $\boldsymbol{\Kappa}$ matrices are used.
All linear regression (LR) models are trained using Scikit-Learn. \cite{pedregosa_scikit-learn_2011}
All Gaussian process regression (GPR) \cite{rasmussen_gaussian_2006} models use the Matern 5/2 kernel \cite{rasmussen_gaussian_2006,genton_classes_nodate} and are optimized using the scaled conjugate gradient option in GPy. \cite{gpy_gpy_2012}
All regression clustering models are trained following the framework outlined in Ref.~\citenum{cheng_regression_2019} using a GPR within each cluster. 

The MOB-ML models for the QM7b-T data set, and the ISO17 data set are trained on non-canonical MP2/cc-pVTZ pair correlation energies. 
Feature selection is performed using random forest regression \cite{breiman_random_2001} with the mean decrease of accuracy criterion, which is sometimes referred to as permutation importance.\cite{breiman_statistical_2001}
All GPR models use the Matern 5/2 kernel and are optimized using the scaled conjugate gradient option in GPy.

\section{Results and Discussion} \label{sec:results}

First, we compare the MOB-ML analytical gradient to the numerical gradient for an exemplary molecule to illustrate the correctness of our derivation and implementation in Table~\ref{tab:analytical_vs_numerical}.
\begin{table}
	\caption{Mean absolute error (MAE) of the MOB-ML analytical nuclear gradient with respect to the MOB-ML numerical nuclear gradient for a non-equilibrium geometry of water. 
	The numerical nuclear gradients in were obtained with a two-step central difference formula with a step size of $5 \times 10^{-4}$ bohr.
	The non-equilibrium geometry of water has bond lengths of 0.986$\text{\AA}$ and 0.958$\text{\AA}$, and a bond angle of 94.5$^{\circ}$. 
	All MOB-ML models are trained on data for 100 water geometries.}
    \label{tab:analytical_vs_numerical}
	\begin{tabular*}{\columnwidth}{p{0.55\columnwidth} c}
	\hline
Regression technique 		& MAE (hartree/bohr) \\
	\hline
	Linear regression      	& $1.45 \times 10^{-8}$ \\
	Gaussian process regression		    & $3.75 \times 10^{-8}$ \\
	Clustered Gaussian process regression     & $2.28 \times 10^{-8}$ \\
	\hline
	\end{tabular*}
\end{table}
Table~\ref{tab:analytical_vs_numerical} shows that the mean absolute errors (MAE) of the analytical MOB-ML gradients of a distorted water molecule with respect to the numerical ones are on the order of $10^{-8}$~hartree/bohr for all MOB-ML models.
A similar MAE is commonly found when comparing analytical and numerical gradients of pure electronic structure methods. \cite{lee_analytic_1991,schutz_analytical_2004,lee_analytical_2019,pinski_analytical_2019}
Additionally, Table~\ref{tab:analytical_vs_numerical} shows that the difference of the numerical and analytical gradient is largely independent of the regression technique (linear regression, Gaussian process regression, or a clustered Gaussian process regression) applied within the MOB-ML model. 
More generally, this illustrates (as also pointed out in Section~\ref{subsubsec:minimizeL}) that any desired regression technique can be applied within MOB-ML without changes to the gradient framework provided that the regression prediction is differentiable with respect to the features.

\begin{figure}
    \includegraphics[width=\columnwidth]{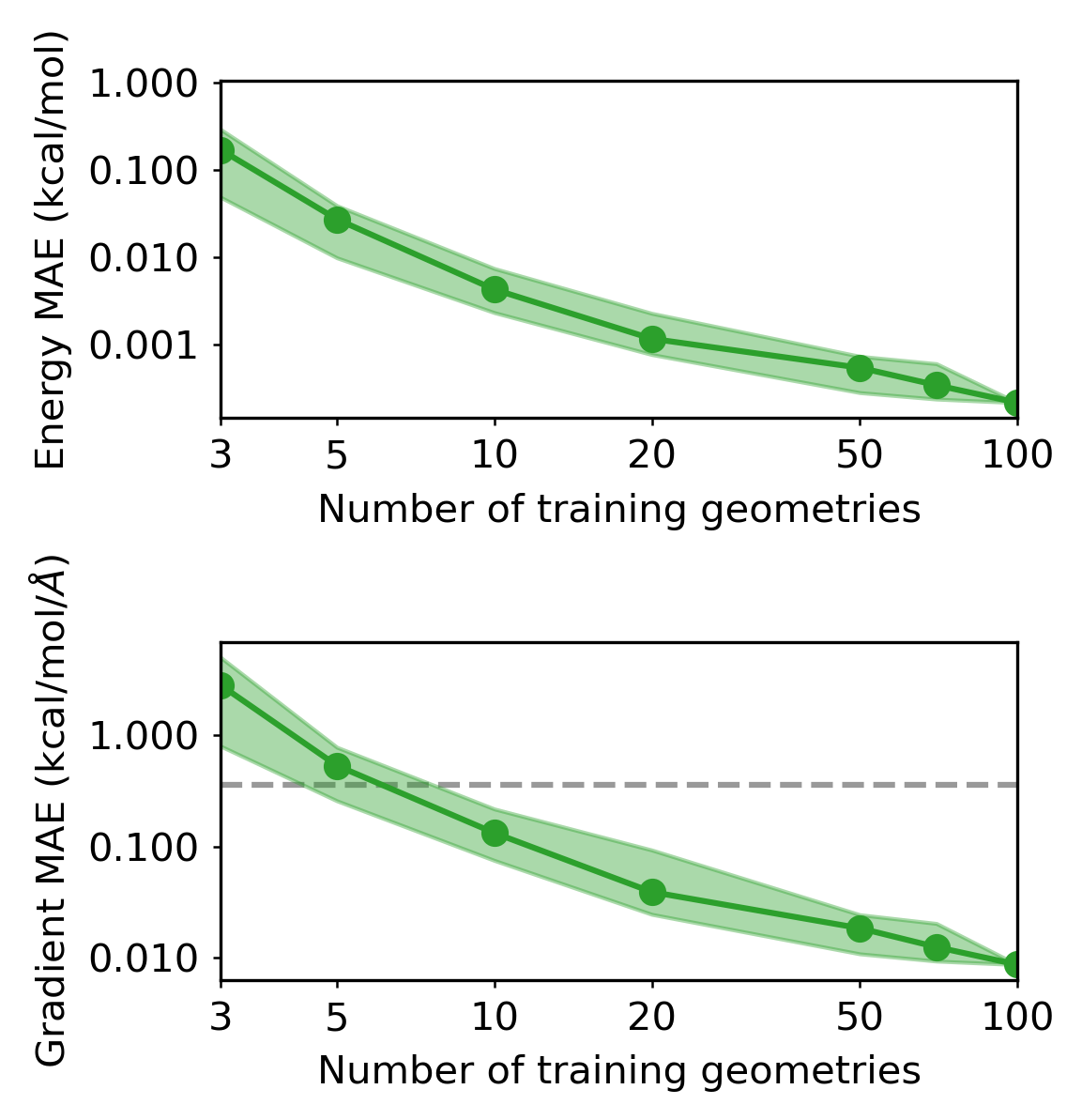}
    \caption{
    MOB-ML learning curves for CCSD(T) energies (top panel) and gradients (bottom panel) for a single water molecule.  Mean absolute errors (MAE) for the predictions are reported as function of the number of water geometries used for training data; only CCSD(T) energies (not gradients) are used for the MOB-ML training data. 
    The green circles correspond to the mean MAE obtained from 50 random samples of the training data, the green shaded area corresponds to the 90\% confidence interval for the predictions  and for the gradients obtained from 50 random samples.
    The black horizontal line at $0.3$~mH/bohr in the bottom panel indicates the commonly used threshold to determine geometry optimization convergence.
    }
    \label{fig:water_learning_curve}
\end{figure}

As a second demonstration, we consider the thermally accessible potential energy surface of a single water molecule, following our previous work. \cite{cheng_universal_2019} 
Fig.~\ref{fig:water_learning_curve} shows the MAE for the energy predictions and for the associated analytical gradients we obtained with MOB-ML models trained on CCSD(T) energies performed on thermalized water 
geometries. 
As already highlighted in Ref.~\citen{cheng_regression_2019}, the MAE for the energy prediction decreases steeply with the number of training geometries and we reach an MAE of $2 \times 10^{-4}$~kcal/mol when training on correlation energies of 100 training geometries. 
Additionally, we see that the MAE of the analytical MOB-ML gradients with respect to the analytical CCSD(T) gradients strictly decreases with an increasing amount of training data although the training data in this context are correlation energy labels and not gradients. 
The MAE of the MOB-ML analytical gradient is $9 \times 10^{-3}$~kcal/mol/\AA\ when training on correlation energies for 100 water geometries.
We can contextualize this result by considering that the threshold commonly used to determine if a structure optimization is converged is $0.36$~kcal/mol/$\text{\AA}$. 
The MAE for the gradient drops below this threshold when training on as few as three to nine water geometries. 
This demonstrates that MOB-ML is able to describe potential energy surfaces to a high accuracy and with a high data efficiency.

\begin{center}
\begin{figure}[htbp]
    \includegraphics[width=\columnwidth]{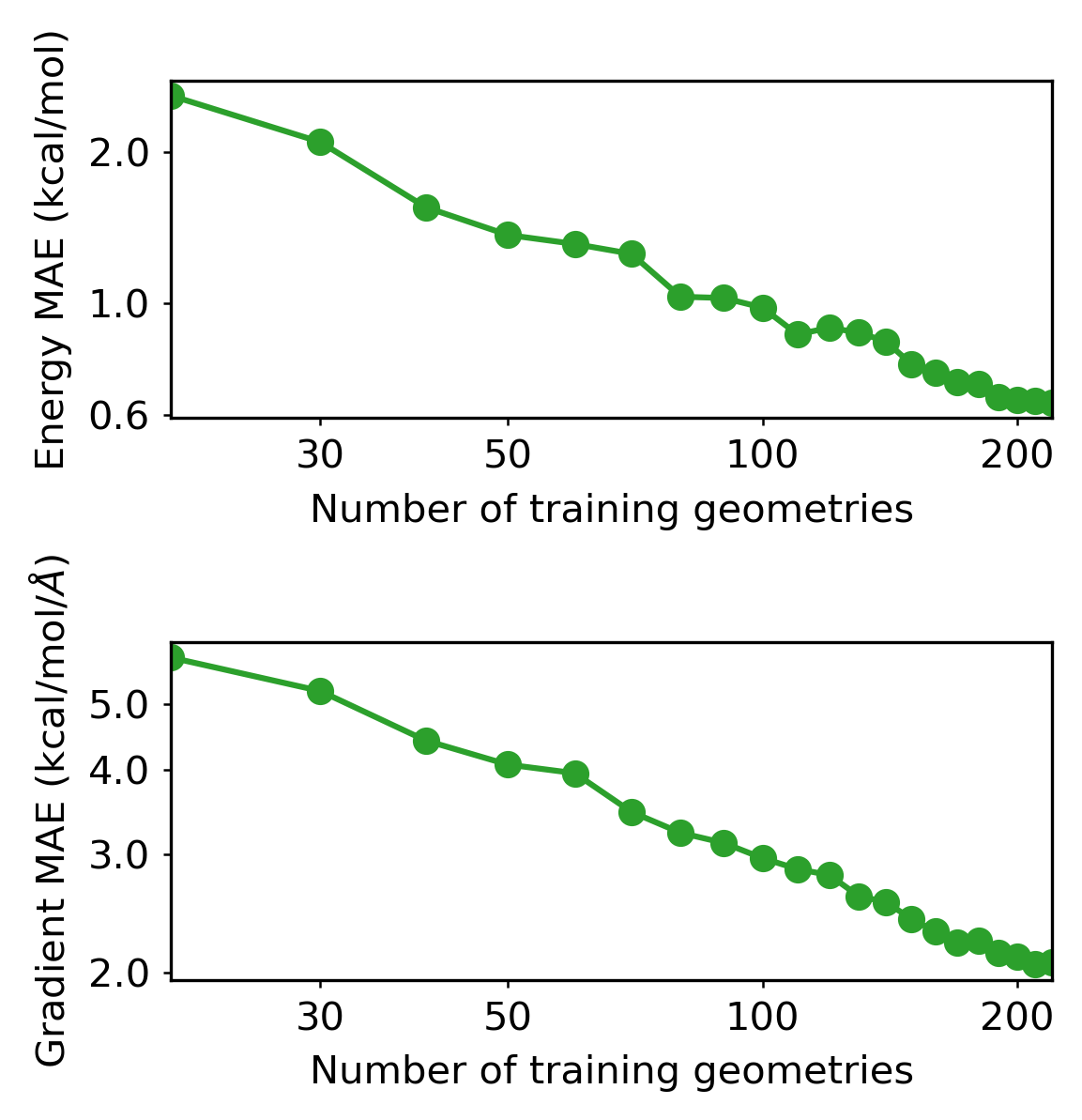}
    \caption{
        MOB-ML learning curves for MP2 energies (top panel) and gradients (bottom panel) for the QM7b-T data set. 
        Mean absolute errors (MAE) for the predictions are reported as function of the number of randomly selected molecules 
        used for training data; only MP2 energies (not gradients) are used for the MOB-ML training data.} 
    \label{fig:qm7b_learning}
\end{figure}
\end{center}

In Fig.~\ref{fig:qm7b_learning}, we show that this result generalizes to a diverse set of molecules. 
To this end, we first study the QM7b-T data set which is comprised of a thermalized set of 7211 organic molecules with 7 or fewer heavy atoms. \cite{cheng_thermalized_2019}
Fig.~\ref{fig:qm7b_learning} shows the MAE for the MOB-ML energy prediction and for the associated analytical gradient with respect to the corresponding MP2 quantities as a function of the number of MP2 reference energy calculations. 
As already reported in Ref.~\citen{husch_enforcing_2020}, the learning curve for the energy decreases steeply and we obtain an MAE of 1.0 kcal/mol when training on about 70 structures.
The decrease in the MAE for the energy prediction is accompanied by a decrease in the MAE for the analytical MOB-ML gradient with respect to the analytical MP2 gradient. 
We reach a MAE of $2.08$ kcal/mol/\AA~when training on 220 structures. 

To compare MOB-ML for gradient predictions with other machine learning methods, we now also examine the ISO17 data set \cite{schutt_schnet_2017}.
The ISO17 data set consists of conformers taken from MD trajectories for constitutional isomers with the chemical formula C$_7$O$_2$H$_{10}$. 
Table~\ref{tab:iso17} shows the performance of two MOB-ML models, one trained on 220 QM7b-T structures and one trained on 100 ISO17 structures, and summarizes the MAEs obtained with other ML models in the literature, i.e., SchNet, \cite{schutt_schnet_2017} FCHL\cite{christensen_operators_2019},  PhysNet \cite{unke_physnet_2019}, the shared-weight neural network (SWNN) \cite{profitt_shared-weight_2019}, GM-sNN, \cite{zaverkin_gaussian_2020} and GNNFF. \cite{park_accurate_nodate}
The MOB-ML models are the only ML models which are on average chemically accurate although the MOB-ML models were only trained on energies for 100 ISO17 molecules and 220 QM7b-T molecules, respectively.
The fact that our model trained on a small set of the seven-heavy atom molecules which are smaller in size than ISO17 and which are chemically more diverse (QM7b-T additionally contains the elements N, S, Cl) showcases again how transferable and data efficient MOB-ML models are. 
The next best model in terms of the energy MAE is GM-sNN which was trained on energies and gradients for 400k ISO17 structures and achieves an MAE of 1.97 kcal/mol.
The force MAE of the MOB-ML models (1.63 and 1.64 kcal/mol/\AA, respectively) is comparable to that of GM-sNN (1.66 kcal/mol/\AA) while employing only 0.025\% of the training data. 
MOB-ML is significantly more accurate in the forces than other models trained on energies alone, i.e., SchNet which obtained an MAE of 5.71 kcal/mol/\AA~and SWNN which obtained an MAE of 6.61 kcal/mol/\AA. 
The only model which is more accurate in terms of the force MAE is PhysNet which is trained on energies and forces for 400k ISO17 structures. 
PhysNet obtains a force MAE of 1.38 kcal/mol/A.  
Given the demonstrated learnability of forces, it is very likely that MOB-ML could be trained to be more accurate by including more training data. 
Furthermore, analytical gradients have not been derived for all reference theories which considerably limits the scope of these machine learning methodologies. 
For example, the popular local coupled cluster methods\cite{schwilk_scalable_2017, guo_communication_2018, nagy_optimization_2018} do not currently have derived analytical gradient theories.
\onecolumngrid
\begin{center}
\begin{table}[htbp]
	\caption{Comparison of the mean absolute error for the prediction of energies and atomic forces for the unknown test set of the ISO17 data set obtained with different ML methods. The different ML methods applied different training sizes and drew on different labels to train the models on.
	Energy and force errors are reported in kcal/mol and kcal/mol/$\text{\AA}$, respectively. }
    \label{tab:iso17}
	\begin{tabular*}{\columnwidth}{p{0.155\columnwidth} p{0.155\columnwidth}  p{0.155\columnwidth} p{0.155\columnwidth}p{0.155\columnwidth} p{0.155\columnwidth}}
	\hline
	Method 		& Training Size & \multicolumn{2}{c}{Trained on energy labels} & \multicolumn{2}{c}{Trained on energy+gradient labels} \\ 
	            &               & Energy MAE & Force MAE & Energy MAE & Force MAE \\
	\hline
	SchNet\cite{schutt_schnet_2017}       & 400,000 & 3.11 & 5.71 & 2.40 & 2.18 \\
	FCHL\cite{christensen_operators_2019} & 1,000   & ---  & ---  & 3.70  & 3.50 \\
	PhysNet \cite{unke_physnet_2019} &  400,000 & ---  & ---  & 2.94 & 1.38 \\
	SWNN \cite{profitt_shared-weight_2019} & 400,000 & 3.72 & 6.61 & 8.57 & 6.74\\
	GM-sNN\cite{zaverkin_gaussian_2020} & 400,000 & ---  & ---  & 1.97 & 1.66 \\
	GNNFF \cite{park_accurate_nodate} & 400,000 & ---  & --- & --- & 2.02 \\ 
 	\textbf{MOB-ML}	    & \textbf{100}        & \textbf{0.84} & \textbf{1.64}  & --- & --- \\
	\textbf{MOB-ML}	    & \textbf{220$^*$}    & \textbf{0.76} & \textbf{1.63}  & --- & ---\\
	\hline
	\end{tabular*}
	$^*$This MOB-ML model was trained on 220 randomly selected structures from the QM7b-T data set.
\end{table}
\end{center}
\twocolumngrid

\begin{center}
\begin{figure}[htbp]
    \includegraphics[width=\columnwidth]{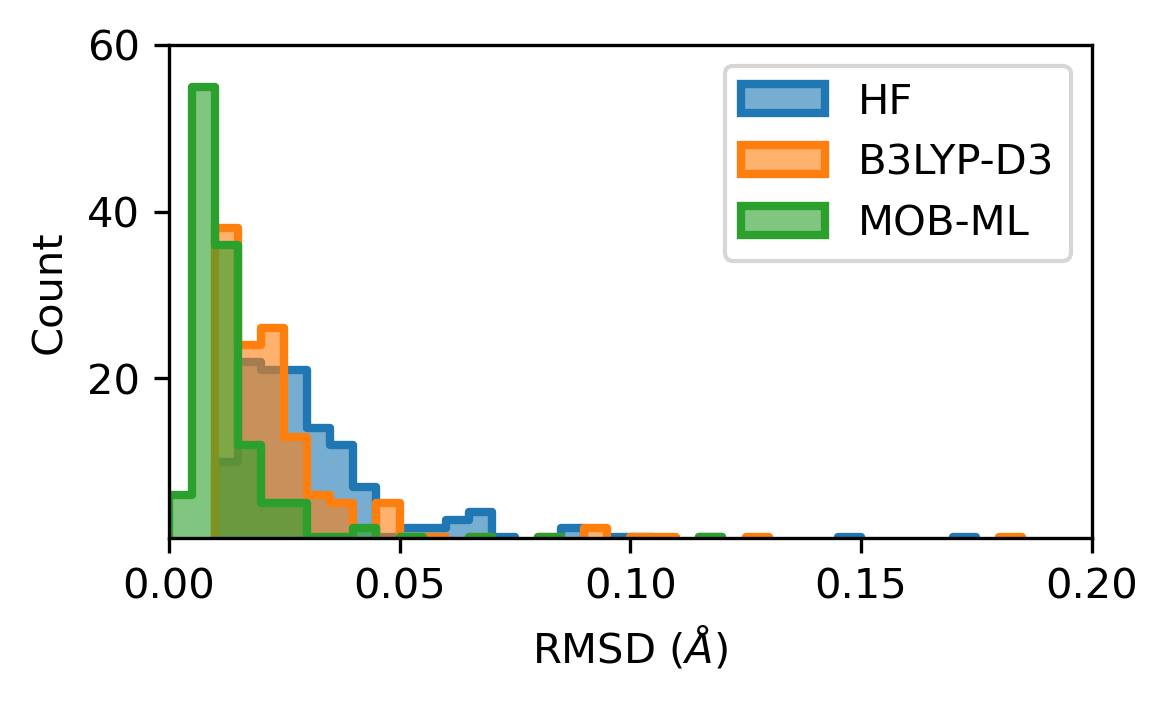}
    \caption{Histogrammed root mean square deviations (RMSD) of HF structures (blue), B3LYP-D3 structures (orange), and MOB-ML structures (green) with respect to MP2 structures for the unique isomers in the ISO17 data set.
    The MOB-ML was trained on 220 randomly selected QM7b-T structures.
    }
    \label{fig:iso17_rmsd}
\end{figure}
\end{center}

Despite comparing favorably to other ML methods, it remains to be shown if the MOB-ML gradients are sufficiently accurate for practical applications. 
Therefore, we now use the MOB-ML gradients to perform the common quantum-chemical task of optimizing molecular structures. 
We optimize the constitutional isomers in ISO17 with MP2 and with MOB-ML and compare the resulting structures via the root mean square deviation (RMSD) of the atoms positions in Figure~\ref{fig:iso17_rmsd}.
Figure~\ref{fig:iso17_rmsd} shows that the MOB-ML optimized structures are very similar to the reference MP2 optimized structures with a mean RMSD of 0.01~\AA. 
The MOB-ML optimized structures are significantly and systematically closer to the reference MP2 structures than the HF-optimized structures which exhibit an average RMSD of 0.03~\AA. 
Moreover, the MOB-ML structures are more similar to the reference MP2 structures than those obtained from B3LYP-D3, a typical DFT exchange-correlation functional. 
The B3LYP-D3 structures exhibit an average RMSD of 0.03~\AA~with respect to the MP2 reference structures. 

\section{Conclusions}

In this work, we have presented the derivation and implementation of the formally complete MOB-ML analytical nuclear gradient theory within a general Lagrangian framework.
We have validated our derivation and implementation by comparison of numerical and analytical gradients.
The MOB-ML gradient framework can be applied in conjunction with any desired fitting technique (e.g., Gaussian process regression or neural networks) and any desired recipe for assembling the MOB-ML feature information.
Furthermore, the framework for evaluating the gradient of a predicted high-accuracy wave function energy is independent of the wave function method MOB-ML was trained to predict. 
Hence, we can take the analytical gradient of a MOB-ML method trained to predict an arbitrary accurate wave function theory.

MOB-ML was previously shown to predict high-accuracy wave function energies at the cost of a molecular orbital evaluation. 
We now have shown that a MOB-ML model trained on correlation energies alone also yields highly accurate gradients for potential energy surfaces of a single molecule and for sets of diverse molecules. 
Specifically, we presented a MOB-ML model which obtains a force MAE of 1.64 kcal/mol/\AA\ for the ISO17 set when only trained on reference energies for 100 molecules beating out the next best model only trained on energies in the literature, SchNet (5.71 kcal/mol/\AA) which was trained on 400k molecules \cite{schutt_schnet_2017}. 
The transferability and data efficiency becomes even clearer when considering that we obtain an MAE of 1.63  kcal/mol/\AA\ for the ISO17 set when training on 220 QM7b-T molecules which are smaller in size (seven versus nine heavy atoms) and which are more diverse in terms of chemical composition. 
The accuracy of a MOB-ML model trained on energies for 220 QM7b-T molecules for the forces is on par with some of the best ML models trained on energies and forces for hundreds of thousands of ISO17 molecules.
Furthermore, we have demonstrated that a force MAE of this magnitude translates into structures which are very close to reference structures. 
Specifically, we obtain a mean RMSD of 0.01~\AA\ with respect to MP2 optimized structures for the ISO17 data set which is is three times smaller than for HF or B3LYP-D3 optimized structures. 
Natural objectives for future work include (i) the inclusion of gradients in the training process to boost the performance in the very low data regime; (ii) the extension to an open-shell framework; (iii) the adaptation of the  Lagrangian framework to derive the analytical gradients of the MOB-ML energy with respect to quantities such as electric and magnetic fields.

\begin{acknowledgments}
This work is supported in part by the U.S. Army Research Laboratory (W911NF-12-2-0023), the U.S. Department of Energy (DE-SC0019390), the Caltech DeLogi Fund, and the Camille and Henry Dreyfus Foundation (Award ML-20-196). S.J.R.L. thanks the Molecular Software Sciences Institute (MolSSI) for a MolSSI investment fellowship. T.H. acknowledges funding through an Early Post- Doc Mobility Fellowship by the Swiss National Science Foundation (Award P2EZP2\_184234).  Computational resources were provided by the National Energy Research Scientific Computing Center (NERSC), a DOE Office of Science User Facility supported by the DOE Office of Science under contract DE- AC02-05CH11231.

\end{acknowledgments}

\section*{Supplementary Material}
The Supplementary Material contains the partial derivatives of the feature vector elements.

\section*{DATA AVAILABILITY STATEMENT}
The data that supports the findings of this study are available within the article and its supplementary material.
The data set used in Table~\ref{tab:analytical_vs_numerical} and Fig.~\ref{fig:water_learning_curve} is available from Ref.~\citenum{cheng_thermalized_2019}.
The data set used in Fig.~\ref{fig:qm7b_learning} is available from Ref.~\citenum{cheng_thermalized_2019}.
The data set used in Table~\ref{tab:iso17} and Fig.~\ref{fig:iso17_rmsd} is available from Ref.~\citenum{schutt_schnet_2017}.

\appendix

\section{Foster-Boys Localization}
\label{appendix:Boys}
This appendix provides additional details for the Boys-related terms in Eqs.~\ref{eq:azloc}, \ref{eq:zvalocc_cpl} and \ref{eq:Wxyz} of the main text. 
The localization conditions for Foster-Boys are \cite{pinski_analytical_2019}
\begin{equation}
    r_{ij} = \sum_{n} R^{n}_{ij} \paran*{R^{n}_{ii} - R^{n}_{jj} } = 0 \text{ for all } i>j \text{,}
\end{equation}
where $n$ corresponds to the x, y, z-coordinates of the position operator. The matrices $R^{n}$ are defined as
\begin{equation}
    R^{n}_{ij} = \sum_{\mu} (i | n | j ) \text{,}
\end{equation}
where $|i)$ and $|j)$ are valence-occupied MOs.
The orbital derivative contributions from the Foster-Boys localization conditions shown in Eq.~\ref{eq:azloc} are
\begin{equation}
\begin{split}
\paran*{ \azvalocc }_{p q} & = \sum_{i>j} \mathcal{B}_{pqij} z_{i j}^{\text{loc}} \Big|_{q \in \text{val-occ}}\text{,}
\end{split}
\end{equation}
where 
\begin{equation}
\begin{split}
    \mathcal{B}_{pqkl} & = \sum_{n} \Big[ \Big( 2 R^{n}_{p k} \delta_{k q} - 2 R^{n}_{p l} \delta_{l q} \Big) R^{n}_{k l} \\
    & + \Big( R^{n}_{k k} - R^{n}_{l l} \Big) \Big( R^{n}_{p l} \delta_{k q} + R^{n}_{p k} \delta_{l q} \Big) \Big] \text{.}
\end{split}
\end{equation}
Next, the dipole derivative contribution from the localization conditions from the first term on the RHS of Eq.~\ref{eq:Wxyz} is
\begin{equation}
\begin{split}
    \paran*{\mathbf{W}_n}_{\mu \nu} &= \tfrac{1}{2} \sum_{ij} C_{\mu i} C_{\nu j} z^{\text{val-occ}}_{ij} \paran*{R^{n}_{ii} - R^{n}_{jj}} \\
    & + \sum_{i} C_{\mu i} C_{\nu i} \sum_{j} z^{\text{val-occ}}_{ij} R^{n}_{ij} \text{.}
\end{split}
\end{equation}
For a full derivation of the orbital and dipole derivatives of Foster-Boys localization conditions please refer to Ref.~\citenum{pinski_analytical_2019}.

\section{IBO Localization}
\label{appendix:IBO}
This appendix provides additional details for the terms IBO-related terms in Eqs.~\ref{eq:ciao_vir}, \ref{eq:azvalvir}, \ref{eq:zcpl_valvir} and \ref{eq:X1} of the main text. 
The localization conditions for IBO are \cite{dornbach_analytical_2019}
\begin{equation}
\begin{split}
    r_{ab} &= 4 \sum_{A} Q^{A}_{ab} \paran*{ \paran*{Q^{A}_{aa}}^{3} - \paran*{Q^{A}_{bb}}^{3} } \\
    & = 0 \text{ for all } a>b \text{,}
\end{split}
\end{equation}
where $A$ corresponds to an atom in the molecule. 
The matrices $Q^{A}$ are defined as 
\begin{equation}
    Q^{A}_{ab} = \sum_{\mu \in A} C^{\text{IAO}}_{\mu a} C^{\text{IAO}}_{\mu b} \text{,}
\end{equation}
where the summation over $\mu$ is restricted to basis functions at atom $A$. The matrix $\mathbf{C}^{\text{IAO}}$ is the MO coefficient matrix represented in the intrinsic atomic orbital (IAO) basis which is defined as
\begin{equation} \label{eq:ciao}
    \mathbf{C}^{\text{IAO}} = \mathbf{X}^{\text{IAO,}\dagger} \mathbf{S}_1 \mathbf{C} \text{,}
\end{equation}
where $\mathbf{S}_1$ is the overlap matrix in the original atomic orbital (AO) basis and $\mathbf{C}$ is the MO coefficient matrix in the original AO basis. 
The matrix that transforms from the AOs to the IAOs shown in Eqs.~\ref{eq:ciao} and \ref{eq:ciao_vir} is 
\begin{equation} \label{eq:xiao}
    \mathbf{X}^{\text{IAO}} = \mathbf{\bar{X}}^{\text{IAO}} \paran*{ \mathbf{\bar{X}}^{\text{IAO},\dagger} \mathbf{S}_1 \mathbf{\bar{X}}^{\text{IAO}} }^{-1/2}
\end{equation}
where
\begin{equation}
    \mathbf{\bar{X}}^{\text{IAO}} = \paran*{ \mathbf{S}_1^{-1} + \mathbf{L} \mathbf{L}^\dagger - \mathbf{\tilde{L}} \mathbf{\tilde{L}}^\dagger } \mathbf{S}_{12} \text{.}
\end{equation}
The matrix $\mathbf{L}$ is the subset of the MO coefficient matrix being localized.
The matrix $\mathbf{\tilde{L}}$ is 
\begin{equation}
    \mathbf{\tilde{L}} = \mathbf{\bar{L}} \paran*{ \mathbf{\bar{L}}^\dagger \mathbf{S}_1 \mathbf{\bar{L}} }^{-1/2}
\end{equation}
where
\begin{equation} \label{eq:L_bar}
    \mathbf{\bar{L}} = \mathbf{S}_1^{-1} \mathbf{S}_{12} \mathbf{S}_{2}^{-1} \mathbf{S}_{12}^\dagger \mathbf{L} \text{.}
\end{equation}
The orbital derivative contributions from the IBO localization conditions shown in Eq.~\ref{eq:azvalvir} corresponds to Eq.~60 in Ref.~\citenum{dornbach_analytical_2019}.
The tensor $\mathcal{C}$ from Eq.~\ref{eq:zcpl_valvir} corresponds to Eq.~37 in Ref.~\citenum{dornbach_analytical_2019}.
The overlap derivative contributions from the IBO localization conditions shown in Eq.~\ref{eq:X1} correspond to Eqs.~50~-~52 in Ref.~\citenum{dornbach_analytical_2019}.

\section{Valence Virtual Conditions}
\label{appendix:valence_virtual}
This appendix provides additional details for the terms in Eqs.~\ref{eq:ciao_vir} and \ref{eq:X1} of the main text. 
In Eq.~\ref{eq:ciao_vir} the matrix $\mathbf{X}_{\text{occ}}^{\text{IAO}}$ is calculated using Eqs.~\ref{eq:xiao}~-~\ref{eq:L_bar} where the matrix $\mathbf{L}$ corresponds to all occupied MOs. 
Next, the overlap derivative contributions from the valence virtual conditions in Eq.~\ref{eq:X1} is
\begin{equation}
\begin{split}
    \sum_{w a} & \frac{ \partial P_{wa} }{ \partial \mathbf{S}_1 } \lambda_{wa} = \mathbf{X}^{\text{IAO}}_{\text{occ}} \mathbf{X}^{\text{IAO},\dagger}_{\text{occ}} \mathbf{S}_1 \mathbf{C}_{\text{vv}} \boldsymbol{\lambda}^\dagger \mathbf{C}_{\text{nvv}}^\dagger \\
    & + \paran*{ \mathbf{X}^{\text{IAO}}_{\text{occ}} \mathbf{X}^{\text{IAO},\dagger}_{\text{occ}} \mathbf{S}_1 \mathbf{C}_{\text{vv}} \boldsymbol{\lambda}^\dagger \mathbf{C}_{\text{nvv}}^\dagger }^\dagger \\
    & + \tfrac{1}{2} \paran*{ \mathbf{\tilde L} \mathbf{\tilde L}^\dagger \mathbf{\bar X} \mathbf{\tilde L} \mathbf{\tilde L}^\dagger - \mathbf{S}^{-1}_{1} \mathbf{\bar X} \mathbf{S}^{-1}_{1} - \mathbf{\tilde X} } \\
    & - \mathbf{\bar X}^{\text{IAO}} \mathbf{H} \mathbf{\bar X}^{\text{IAO},\dagger} \text{.}
\end{split}
\end{equation}
The matrices $\mathbf{\bar X}$, $\mathbf{\tilde X}$, and $\mathbf{H}$ are the same as the Eqs.~56, 57 and 48, respectively, shown in Ref.~\citenum{dornbach_analytical_2019}. 
The evaluation of these matrices differ here by redefining the matrix $\mathbf{G}$ (Eq.~42 in Ref.~\citenum{dornbach_analytical_2019}), to be
\begin{equation}
    \mathbf{G} = \mathbf{S}_1 \mathbf{C}_{\text{nvv}} \boldsymbol{\lambda} \mathbf{C}_{\text{vv}}^\dagger \mathbf{S}_{1} \mathbf{X}^{\text{IAO}}_{\text{occ}} \text{,}
\end{equation}
the matrix $\mathbf{B}$ to be
\begin{equation}
    \mathbf{B} = \mathbf{X}^{\text{IAO}}_{\text{occ}} \text{,}
\end{equation}
and the matrix $\mathbf{L}$ to span all occupied MOs.
The overlap derivative contributions from the valence virtual conditions in Eq.~\ref{eq:X1} are
\begin{equation}
    \sum_{w a} \frac{ \partial P_{wa} }{ \partial \mathbf{S}_2 } \lambda_{wa} = - \tfrac{1}{2} \mathbf{S}^{-1}_2 \mathbf{S}^\dagger_{12} \mathbf{\check X} \mathbf{S}_{12} \mathbf{S}^{-1}_2 \text{,}
\end{equation}
and
\begin{equation}
\begin{split}
    \sum_{w a} \frac{ \partial P_{wa} }{ \partial \mathbf{S}_{12} } \lambda_{wa} & = \paran*{ \mathbf{S}^{-1}_1 + \mathbf{L} \mathbf{L}^\dagger - \mathbf{\tilde L} \mathbf{\tilde L}^\dagger } \mathbf{V} \\
    & \quad + \mathbf{ \check X } \mathbf{S}_{12} \mathbf{S}^{-1}_{2} \text{.}
\end{split}
\end{equation}
The matrices $\mathbf{\check X}$, and $\mathbf{V}$ are evaluated by Eqs.~58 and 54, respectively, in Ref.~\citenum{dornbach_analytical_2019} with the same modifications to $\mathbf{G}$, $\mathbf{B}$ and $\mathbf{L}$.

\section{Density Fitting Approximation}
This appendix provides details on how the density fitting approximation can be used to approximate the four-center AO integral derivatives in Eq.~\ref{eq:mobml_grad_ao}.
The AO integral derivatives are approximated by
\begin{equation} \label{eq:df_approx}
\begin{split}
    (\mu \nu | & \kappa \sigma)^{(q)} \approx \paran*{\mu \nu | \kappa \sigma}^{(q)}_{\text{DF}} = \sum_{P} \paran*{\mu \nu | P}^{(q)} c^{P}_{\kappa \sigma} \\
    & + \sum_{P} c^{P}_{\mu \nu} \paran*{P | \kappa \sigma}^{(q)} - \sum_{PQ} c^{P}_{\mu \nu} J^{(q)}_{PQ} c^{Q}_{\kappa \sigma} \text{,}
\end{split}
\end{equation}
where $P$ and $Q$ label density fitting basis functions, $\paran*{\mu \nu | P}^{(q)}$ are three-center AO integrals, and $J_{PQ}$ are two-center AO integrals. The matrix $\mathbf{c}^{P}$ is
\begin{equation}
    c^{P}_{\kappa \sigma} = \sum_{Q} \sqbrak*{\mathbf{J}^{-1}}_{PQ} \paran*{Q|\kappa \sigma} \text{.}
\end{equation}
Substituting Eq.~\ref{eq:df_approx} into Eq.~\ref{eq:mobml_grad_ao} yields
\begin{equation}
\begin{split}
  \sum_{\mu \nu \kappa \sigma} & \paran*{\mu \nu | \kappa \sigma}^{(q)}_{\text{DF}} \sum_{pq} D_{\mu \nu \kappa \sigma} =\\
  & 2 \sum_{P \mu \nu} \paran*{\mu \nu | P}^{(q)} \Lambda^{P}_{\mu \nu} - \sum_{PQ} J^{(q)}_{PQ} \Gamma_{PQ} \\ 
\end{split}
\end{equation}
where 
\begin{equation}
    c^{P}_{pq} = \sum_{Q} \sqbrak*{\mathbf{J}^{-1}}_{PQ} \paran*{Q|pq} \text{,}
\end{equation}
\begin{equation}
\begin{split}
    \Lambda^P_{\mu \nu} & = \paran*{ \mathbf{d}_{\text{b}} }_{\mu \nu} \sum_{\kappa \sigma} \gamma_{\sigma \kappa}  c^P_{\sigma \kappa} \\
    & - \tfrac{1}{2} \sum_{\kappa \sigma}  \gamma_{\sigma \mu} \paran*{ \mathbf{d}_{\text{b}} }_{\mu \kappa} c^P_{\sigma \kappa}\\
    & + 2 \sum_{q} C_{\mu q} C_{\nu q} \sum_{p} c^{P}_{pp} \denJ_{pq} \\
    & + 2 \sum_{q} C_{\nu q} \sum_{p} c^{P}_{pq} \denK_{pq} C_{\mu p} \text{,} 
\end{split}
\end{equation}
and 
\begin{equation}
\begin{split}
    \Gamma_{PQ} & = \sum_{\mu \nu} \paran*{\mathbf{d}_{\text{b}}}_{\mu \nu} c^{P}_{\mu \nu} \sum_{\kappa \sigma} \gamma_{\sigma \kappa} c^{Q}_{\sigma \kappa} \\
    & - \tfrac{1}{2} \sum_{\mu \nu \kappa \sigma} \paran*{\mathbf{d}_{\text{b}}}_{\mu \kappa} \gamma_{\nu \sigma} c^P_{\mu \nu} c^Q_{\kappa \sigma} \\
    & + 2 \sum_{pq} c^{P}_{pp} c^{Q}_{qq} \denJ_{pq} \\
    & + 2 \sum_{pq} c^{P}_{pq} c^{Q}_{pq} \denK_{pq} \text{.}
\end{split}
\end{equation}

\bibliography{pruned_mob_ml_gradient}

\end{document}